%Paper: hep-th/9407081
%From: C H Oh <PHYOHCH%NUSVM.bitnet@CUNYVM.CUNY.EDU>
%Date: Fri, 15 Jul 94 16:07:16 SST
%Date (revised): Thu, 21 Jul 94 17:37:07 SST

%PlainTex
\raggedbottom
\hoffset=0cm\voffset=0cm
\magnification=\magstep1
\baselineskip=15pt
\def\itm#1{\noindent\item{(#1)}}
\def\ref#1{\noindent\item{[#1.]}}
\def\spa#1{\noindent\item {#1}}
\vskip 9cm
\hfill {\tenrm NUS/HEP/92200}
\par\noindent
\hfill {\tenrm hep-th/9407081}
\centerline {\bf Conformal Properties of Primary Fields in a
$q$-Deformed Theory}
\vskip .7cm
{\centerline {C.H. Oh ${}^{\dag}$ and K. Singh ${}^{\ddag}$}}
\vskip .5cm
{\sl {\centerline {Department of Physics}}}
{\sl {\centerline {Faculty of Science}}}
{\sl {\centerline {National University of Singapore}}}
{\sl {\centerline {Lower Kent Ridge, Singapore 0511}}}
{\sl {\centerline {Republic of Singapore}}}
\vskip 2cm
{\bf {\centerline {Abstract}}}
\vskip .5cm
\noindent
We examine some of the standard features of primary fields in the
framework of a $q$-deformed conformal field theory. By introducing
a $q$-OPE between the energy momentum tensor and a primary field,
we derive the $q$-analog of the conformal Ward identities for
correlation functions of primary fields. We also obtain solutions
to these identities for the two-point function.
\vskip 6truecm
\centerline{Published in J.Phys. A: Math. Gen. (1992) {\bf 25} L149}
\vskip.3cm
\hrule width6cm
\vskip .3truecm
\sevenrm \dag E-mail: PHYOHCH@NUSVM.Bitnet
\vskip .1truecm
\ddag E-mail: PHYSINGH@NUSVM.Bitnet
\vfil
\eject
\par
In recent years, there has been growing interest in the study of
quantized universal enveloping algebras. Loosely called quantum
groups, they first appeared in the study of the Quantum Yang-Baxter
Equations related to the inverse scattering problem [1].
Subsequently, it was shown that they can be obtained from
representations of mathematical structures called quasi-triangular
Hopf algebras [2]. These structures which often depend on a
parameter $q$ can be regarded as $q$-deformations of Lie algebras
in the sense that as $q\to 1$ the algebra reduces to the usual Lie algebra.
\vskip .3cm
\par
Explicit realizations of some of these quantum groups have been
obtained by many authors [2-4]. For instance the Jordan-Schwinger
approach often used in the study of angular momentum algebra has
been suitably generalized to give bosonic ($q$-oscillator)
representations of the quantum group ${\rm SU}_{q}(2)$ [4].
More recently, Curtright and Zachos [5] have constructed a
$q$-analoque of the centreless Virasoro algebra by using
a differential realization of ${\rm SU}_{q}(1,1)$ (see also Ref.[6]). The
central extension to this algebra has been furnished by Aizawa and
Sato [7]. In fact, they have also found a $q$-deformed operator
product expansion (OPE) between two energy momentum tensors which
realizes this algebra. This naturally paves the way for a
$q$-deformed conformal field theory.
\vskip .3cm
\par
In this letter we study the properties of primary fields in the
spirit of Ref.[7]. Here we reexamine some of the well known issues
pertaining to standard conformal field theory (CFT) [8] in the context
of such a $q$-deformed theory. In particular, we introduce a
$q$-OPE of the energy momentum tensor with a primary field which
extends the $q$-OPE of Ref.[7] to primary fields of arbitrary
conformal weights. The deformation reflected in this $q$-OPE is
shown to be equivalent to the one used by Chaichan et.al. [9].
Using arguments paralleling those used in standard CFT, we obtain the
$q$- analog of the conformal Ward identity and
the projective Ward identities for correlation functions of
primary fields. In particular, for the two-point function, it is
shown that these Ward identities do not uniquly determine it when
$\vert q \vert =1$. Using
the $q$-OPE we also realize the algebra between the modes of the
energy momentum tensor and those of a holomorphic primary field.
\vskip .3cm
\par
We begin by summarizing some basic features of standard CFTs [8,10]
that will be used or modified later. Consider a primary field
$\Phi(z,\overline z)$ with conformal weights $h, \overline h$. It
is defined by its transformation under $z\to z'=f(z),\overline z
\to \overline z'= \overline f(\overline z)$:
$$\Phi(z,\overline z)\to \Phi '(z,\overline z)
= (\partial f)^h (\overline \partial \overline f)^{\overline h}\Phi
(f(z),\overline f (\overline z))\eqno(1)$$
where $f(z)$ and $\overline f(\overline z)$ are arbitrary
holomorphic and antiholomorphic functions respectively. When the
transformation is infinitisimal, i.e., $f(z)=z + \epsilon (z)$ and
$\overline f (\overline z) = \overline z + \overline \epsilon (\overline z)$,
then
$$\Phi '(z,\overline z)=\Phi (z,\overline z) +
\Delta _{\epsilon,\overline \epsilon}\Phi(z,\overline z)\eqno(2a)$$
with
$$\Delta _{\epsilon,\overline \epsilon}\Phi(z,\overline z)=
(h\partial \epsilon + \epsilon \partial)\Phi(z,\overline z) +
(\overline h \overline \partial \overline\epsilon + \overline \epsilon
\overline \partial)\Phi(z,\overline z)\eqno(2b)$$
In particular when $\epsilon (z)=\epsilon_n z^{n+1}$ and
$\overline \epsilon (\overline z)=\overline \epsilon_n \overline
z^{n+1}$ where $\epsilon_n$ and $\overline \epsilon_n$ are small
constants, we have
$$\Delta_{n}\Phi (z,\overline z) = \epsilon_n \delta _n \Phi
(z,\overline z)+\overline \epsilon_m \overline \delta _m \Phi
(z,\overline z)\eqno(3a)$$
where
$$\delta _n \Phi(z,\overline z)=(z\partial + h(n+1)-n)z^n
\Phi(z,\overline z)\eqno(3b) $$
$$\overline \delta _n \Phi(z,\overline z)=(\overline z\overline\partial +
\overline h(n+1)-n)\overline z^n\Phi(z,\overline z)\eqno(3c) $$
In the following, we will only consider the holomorphic terms with
similar results holding for the antiholomorphic ones.
\vskip .3cm
\par
In a quantum theory, the variation in $\Phi(z,\overline z)$ is
implemented by the "equal-time" commutator:
$$\delta_n \Phi(w,\overline w)=[\oint _{C_{0}}{dz\over {2 \pi i}}z^{n+1}T(z),
\Phi(w,\overline w)]\eqno(4)$$
where $T(z)$ is the holomorphic component of the energy momentum tensor.
On the $z$-plane, different times correspond to concentric circles
of different radii and the notion of time ordering is replaced by
that of radial ordering:
$$R(A(z)B(w))=\cases { A(z)B(w)~~~& $\vert z\vert > \vert w\vert$\cr
B(w)A(z)& $\vert z\vert < \vert w\vert.$\cr}\eqno(5)$$
In this scheme the "equal-time" commutator is given by [10]
$$\eqalign {\oint_{C_{0}} {dz\over {2 \pi i}}z^{n+1}[T(z),\Phi(w,\overline
w)] &= (\oint _{\vert z\vert > \vert w\vert}-
\oint _{\vert z\vert < \vert w\vert}){dz\over {2 \pi i}}z^{n+1}R(T(z)
\Phi(w,\overline w))\cr
&= \oint_{C_{P}}{dz\over {2 \pi i}}z^{n+1}R(T(z)\Phi(w,\overline w))}\eqno(6)$$
where the last integral is taken around all the poles in the OPE of
$T(z)\Phi(w,\overline w)$ which we assume are located on the
$\vert z\vert =\vert w\vert$ contour.  Indeed, by comparison with (3b)
one can infer that
$$T(z) \Phi (w,\overline w)= {{h \Phi (w,\overline w)}
\over {(z-w)^2}}+ {{\partial \Phi (w,\overline w)}\over {(z-w)}}
+ {\rm {regular~terms}}.\eqno(7)$$
\vskip .3cm
\par
Now a $q$-deformation of the theory is achieved by replacing
eqns. (3b) and (3c) by the corresponding $q$-analogs. For this purpose we
consider the deformation as defined by Chaichan et. al.[9]:
$$\delta _n \Phi (z,\overline z)\to \delta_{n}^{q}\Phi (z,\overline
z)=[z\partial + h(n+1)-n]z^n \Phi(z,\overline z)\eqno(8)$$
which essentially replaces the bracket in (3b) by a $q$-bracket
defined by
$$[x]={{q^x-q^{-x}}\over {q-q^{-1}}}.\eqno(9)$$
Eqn.(8) thus, serves as a definition of a primary field in a
$q$-deformed theory.
We will now like to implement this variation as an "equal-time"
commutator as in eqn.(6). To this end, we introduce a
$q$-OPE of $T(z)$ with $\Phi(w,\overline w)$, which we write as,
$$\eqalign{{(T(z)\Phi(w,\overline w))}_q &=
{{[h/2]}\over {(z-w)}}\Bigl\lbrace
{{\Phi (wq^{-1},\overline w)}\over {zq^{h/2}-wq^{-h/2}}}+
{{\Phi (wq,\overline w)}\over {zq^{-h/2}-wq^{h/2}}}\Bigr\rbrace\cr
&+ {1\over {(z-w)}}\partial_{w}^{q}\Phi (w,\overline w)+{\rm {regular~terms}}}
\eqno(10)$$
where $\partial^{q}_{w}$ is the $q$-analog of the derivative:
$$\partial^{q}_{w}f(w)={{f(wq)-f(wq^{-1})}\over{w(q-q^{-1}})}.\eqno(11)$$
Using this definition for the $q$-derivative, we can also rewrite
the $q$-OPE as
$$(T(z)\Phi(w,\overline w))_q =
{{1}\over {w(q-q^{-1})}}\Bigl\lbrace
{{\Phi (wq,\overline w)}\over {z-wq^{h}}}-
{{\Phi (wq^{-1},\overline w)}\over {z-wq^{-h}}}\Bigr\rbrace + {\rm
regular~terms} \eqno(12)$$
which shows that it is singular at the points $z=wq^{\pm h}$.
It is easy to verify that the above OPE leads to the correct
variation in $\Phi$, by evaluating the integral in eqn.(6) with
$C_{P}$ taken as a contour encircling the points $wq^{h}$
and $wq^{-h}$.
Before proceeding further, let us make a few observations:
\itm 1 It is evident from the above expression that there are poles
present at two points rather than one. These poles are both of
order 1, unlike the undeformed case where the $z=w$ pole is of
order 2. It is interesting to note, however, that in the limit $q
\to 1$ these poles will coalesce to form a pole of order 2 at
$z=w$. In fact, in the limit $q\to 1$, our $q$-OPE reduces to the standard
one (eqn.(7)).
\itm 2 Recall that the the "equal-time" commutator in
eqn.(6) was evaluated as a difference of two integrals with
contours which are concentric and close to the $\vert z\vert =\vert
w \vert$ contour
but one having radius $\vert z\vert>
\vert w\vert$ and the other $\vert z\vert<\vert w\vert.$
They combine into a single contour which is taken to be
a small circle centred around the singular point ($z=w$).
For this scheme to be applicable here, we must require that the two
poles lie on the $\vert z\vert =\vert w \vert$ contour, since
otherwise this poles will not make any contributions to the
integral. This means that we must restrict ourselves  to the case
when $\vert q\vert=1$, i.e. $q$ should be taken as a pure phase
$(q=e^{i\alpha \pi})$.
\itm 3 It is worth noting that the variation in $\Phi$ obtained by
our $q$-OPE is similar to the one used by Chaichan et. al. only for
the case $\epsilon (z)=z^{n+1}$. For arbitrary $\epsilon (z)$,
their variation is assumed to be of the form
$$\delta_{\epsilon}^{q}\Phi (z,\overline z)=\epsilon (z)^{1-h}\partial ^{q}_{z}
(\epsilon (z)^{h}\Phi(z,\overline z))\eqno(13)$$
while ours is given by
$$\eqalign {\delta_{\epsilon}^{q}\Phi (z,\overline z) &=
\oint_{C_{P}}{d\xi \over {2 \pi i}} \epsilon (\xi) R(T(\xi)
\Phi(z,\overline z))_{q}\cr
&= \epsilon (zq^{h})\partial ^{q}_{z}\Phi (z,\overline z)+
[h]\partial_{z}^{q^{h}}\epsilon (z)\Phi (zq^{-1},\overline z).}\eqno(14)$$
Both, however, reduces to (8) when $\epsilon (z)$ is taken to be $z^{n+1}$.
\itm 4 When $h=2$, our expression is similar to the one given by
Aizawa and Sato [7] for the OPE of two energy momentum tensors when
the central charge in their expression is taken as zero.
\vskip .3cm
\par
With the $q$-OPE defined as above, we can write down the $q$-Ward identities
for the correlation functions of primary fields.
We begin by considering the action of the generator of
infinitisimal conformal transformations on the correlation of $n$
primary fields $\lbrace \Phi_{i}(w_{i},\overline w_{i})\rbrace$ with
correponding conformal weights $h_{i},\overline h_{i}(i=1,2\ldots n)$:
$$<\oint_{C_{0}}{dz\over {2 \pi i}}\epsilon
(z)T(z)\Phi_{1}(w_{1},\overline
w_{1})\ldots\Phi_{n}(w_{n},\overline w_{n})>_{q}.\eqno(15)$$
(The contour $C_{0}$ encircles all the
points $\lbrace w_{i}q^{hk}\vert k=\pm 1\rbrace_{i=1,2,\ldots n}$.)
In the above expression, the correlation function $<\ldots>_{q}$ is
taken relative to the "in" ($\vert 0>_{q}$) and the "out" (${}_{q}<0 \vert$)
vacuums which are defined by requiring that
$$\eqalignno{L_{m}\vert 0>_{q}&=0~~~~~~~~m\ge -1 &(16a)\cr
{}_{q}<0\vert L_{m}&=0~~~~~~~~m\le 1 &(16b)\cr}$$
where
$$L_{m}=\oint_{C_{0}}{dz\over {2 \pi i}}z^{m+1}T(z),~~~~~~m\in Z,\eqno(17)$$
are modes in the expansion,
$$T(z)=\sum _{m\in Z}L_{m}z^{-m-2}.\eqno(18)$$
Note that conditions (16a) and (16b) ensures the regularity of $T(z)\vert
0>_{q}$ and its adjoint at $z=0$ and $z=\infty$.
By analyticity, the contour $C_{0}$ in expression (15) can be
deformed to a sum of $n$ contours with each contour $C_{i}$ surrounding
the points, $\lbrace w_{i}q^{h},w_{i}q^{-h}\rbrace$.
Then as a consequence of the $q$-OPE, we have
$$\oint_{C_{0}}{dz\over {2 \pi i}}\epsilon
(z)<T(z)\Phi_{1}(w_{1},\overline w_{1})\ldots\Phi_{n}(w_{n},\overline w_{n})>
_{q}~~~~~~~~~~~~~~~~~~~~~~~~~~~~~~~$$
$$\eqalign {&= \sum_{i=1}^{n}<\Phi_{1}(w_{1},\overline w_{1})\ldots
\oint_{C_{i}}{dz\over {2 \pi i}}\epsilon
(z)(T(z)\Phi_{i}(w_{i},\overline w_{i}))_{q}\ldots
\Phi_{n}(w_{n},\overline w_{n})>_{q}\cr
&=\sum_{i=1}^{n}\oint_{C_{i}}{dz\over {2 \pi i}}\epsilon
(z){\cal L}_{z;w_{i}}^{h_{i}}<\Phi_{1}(w_{1},\overline w_{1})\ldots\Phi_{n}
(w_{n},\overline w_{n})>_{q},}\eqno(19)$$
where the differential operator ${\cal L}_{z;w_{i}}^{h_{i}}$ is
given by
$${\cal L}_{z;w_{i}}^{h_{i}}\equiv {{1}\over {(z-w_{i})}}\Bigl\lbrace
[h_{i}/2]\Bigl {(} {{q^{-w_{i}\partial_{w_{i}}}}\over {zq^{h_{i}/2}-w_{i}
q^{-h_{i}/2}}}
+ {{q^{w_{i}\partial_{w_{i}}}}\over {zq^{-h_{i}/2}-w_{i}
q^{h_{i}/2}}}\Bigl {)}+\partial_{w_{i}}^{q}\Bigr \rbrace. \eqno(20)$$
Furthermore, since $\epsilon (z)$ is arbitrary, we can write,
$$<T(z)\Phi_{1}(w_{1},\overline w_{1})\ldots\Phi_{n}(w_{n},\overline w_{n})>
_{q}= \sum_{i=1}^{n}{\cal L}_{z;w_{i}}^{h_{i}}<\Phi_{1}(w_{1},
\overline w_{1})\ldots\Phi_{n} (w_{n},\overline w_{n})>_{q},\eqno(21)$$
which is the unintegrated form of the $q$-Ward identity. It is easy to see
that it reduces to the usual one as $q\to 1$.
\vskip .3cm
\par
Next let us consider the $q$-analog of the projective Ward
Identities. From (16a) and (16b) it is easy to see that the generators
$L_{0,\pm 1}$ annhilate both the "in" and the "out" vacuums. On substituting
$\epsilon (z)=z^{m+1}$ for $m=-1,0,1$ into eqn.(19) and integrating, we have
$$\eqalignno{&\sum_{i=1}^{n}w_{i}^{-1}[w_{i}\partial _{w_{i}}]
<\Phi_{1}(w_{1},
\overline w_{1})\ldots\Phi_{n} (w_{n},\overline w_{n})>_{q}=0 &(22a)\cr
&\sum_{i=1}^{n}[w_{i}\partial _{w_{i}}+h_{i}]
<\Phi_{1}(w_{1},\overline w_{1})\ldots\Phi_{n} (w_{n},\overline w_{n})>_{q}
=0 &(22b)\cr
&\sum_{i=1}^{n}w_{i}[w_{i}\partial _{w_{i}}+2h_{i}]
<\Phi_{1}(w_{1},
\overline w_{1})\ldots\Phi_{n} (w_{n},\overline w_{n})>_{q}=0 &(22c)\cr}$$
for any $n$-point function. These are the $q$-analogs of the
projective Ward identities.
Now, it is well known that in standard
CFT the two-point and three-point functions are severely constrained
by the Ward identities. In fact, they are uniquely determined
up to a normalization constant. The situation for the $q$-deformed
case is not quite the same. Here when $\vert q\vert =1$ the $q$-
Ward identities do not uniquely specify them as we will illustrate
below. For this purpose assume an ansatz for the correlation function of two
primary fields $\Phi_{1}(w_{1},\overline w_{1}),
\Phi_{2}(w_{2},\overline w_{2})$ with conformal weight
$h_{1},h_{2}$ respectively to be of the form
$$<\Phi_{1}(w_{1},\overline w_{1})\Phi_{2}(w_{2},\overline w_{2})>_{q}=
{1\over {(w_{1}-w_{2})^{n}_{q}(\overline w_{1}-\overline
w_{2})^{\overline n}_{q}}}~~~~~~~~~~\vert w_{1}\vert >\vert
w_{2}\vert ,\eqno(23)$$
where
$$(w_{1}-w_{2})^{n}_{q}=\prod_{k=1}^{n}(w_{1}-w_{2}q^{n-2k+1})=
\sum_{k=1}^{n}{{[n]!}\over {[n-k]![k]!}}w_{1}^{n-k}(-w_{2})^{k}\eqno(24)$$
is the $q$-analog of the distance function $(w_{1}-w_{2})^{n}$ [7].
On substitution into (22a), (22b) and (22c) we obtain the following conditions:
$$\eqalignno{&[h_{1}-n]+[h_{2}]=0 &(25a)\cr
&[h_{2}-n]+[h_{1}]=0 &(25b)\cr
&[2h_{1}-n]=0 &(25c)\cr
&[2h_{2}-n]=0 &(25d)\cr
&[2h_{2}]-[2h_{1}]=0.&(25e)}$$
Apart from the obvious solution
$$h_{1}=h_{2}=n/2,\eqno(26)$$
we also have for $q=e^{i\pi \alpha}$,
$$\eqalignno{h_{1} &= n/2 +k/\alpha &(27a)\cr
h_{2} &= n/2 +l/\alpha,&(27b)}$$
where $k$ and $l$ are arbitrary integers which are either both even
or both odd. Adding the two we have
$$n=h_{1}+h_{2}-(k+l)/\alpha \eqno(28)$$
and this means that $n$ which characterizes the solution is not unique
by virtue of the fact that $k$ and $l$ are arbitrary.
\vskip .3cm
\par
It is also interesting to study the commutator algebra (or rather
"quommutator") of the generators $\lbrace L_{n}\rbrace$ with the
modes $\lbrace \phi_{m}\rbrace$ of a primary field. Consider a
holomorphic primary field with conformal weights $(h,0)$,
$$\Phi(w)=\sum_{m\in Z-h}\phi_{m}w^{-m-h},\eqno(29)$$
with the modes $\lbrace \phi_{m}\rbrace$ satisfying
$$\phi_{m}=\oint _{C_{0}}{dw\over {2 \pi i}}w^{m+h-1}\Phi(w).\eqno(30)$$
Here we would like to evaluate the bracket
$$[L_{n},\phi_{m}] \equiv (L_{n}\phi_{m})_{q}-(\phi_{m}L_{n})_{q} \eqno(31)$$
where the terms $(~)_{q}$ are defined via the
$q$-product of two field operators $A(z)$ and $B(w)$ [7]:
$$(A(z)B(w))_{q}\equiv A(zq)B(wq^{-1}).\eqno(32)$$
For instance, we have (following Ref.[7])
$$\eqalign{(L_{n}\phi_{m})_{q}
&= \oint _{C_{1}}{dz\over {2 \pi i}}\oint _{C_{2}}{dw\over {2 \pi i}}
z^{n+1}w^{m+h-1}(T(z)\Phi(w))_{q}\cr
&= \oint _{C_{1}}{dz\over {2 \pi i}}\oint _{C_{2}}{dw\over {2 \pi i}}
z^{n+1}w^{m+h-1}T(zq)\Phi(wq^{-1})\cr
&= \oint _{C_{1}}{dz\over {2 \pi i}}\oint _{C_{2}}{dw\over {2 \pi i}}
z^{n+1}w^{m+h-1}\sum_{k}L_{k}(zq)^{-k-2}\sum_{l}\phi_{l}(wq^{-1})^{-l-h}\cr
&= q^{m-n+h-2}L_{n}\phi_{m}}\eqno(33)$$
where $C_{1}$ and $C_{2}$ are contours about the origin such that
$C_{2}\subset C_{1}$.
Similarly
$$(\phi_{m}L_{n})_{q}=q^{-(m-n+h-2)}\phi_{m}L_{n}.\eqno(34)$$
Then by combining (33) and (34), the bracket in (31) can be reexpressed as
$$[L_{n},\phi_{m}]\equiv q^{m-n+h-2}L_{n}\phi_{m}-
q^{-(m-n+h-2)}\phi_{m}L_{n}.\eqno(35)$$
To evaluate this bracket, we use the $q$-OPE :
$$\eqalign{[L_{n},\phi_{m}]
&=\oint _{C_{0}}{dz\over {2 \pi i}}\oint _{C_{P}}
{dw\over {2 \pi i}}z^{n+1}w^{m+h-1}R(T(z)\Phi(w))_{q}\cr
&=[n(h-1)-m]\phi_{n+m},}\eqno(36)$$
which gives the "quommutator" of $L_{n}$ with $\phi_{m}$,
$$q^{m-n+h-2}L_{n}\phi_{m}-q^{-(m-n+h-2)}\phi_{m}L_{n}=
[n(h-1)-m]\phi_{n+m}.\eqno(37)$$
Again, we can see that this reduces to the standard result as $q \to 1$. It is
also interesting to note that if we identify $\phi_{m}$ with $L_{m}$
with $h=2$ then the above algebra corresponds to the $q$-deformed
centreless Virasoro algebra,
$$q^{m-n}L_{n}L_{m}-q^{-(m-n)}L_{m}L_{n}=
[n-m]L_{n+m}\eqno(38)$$
proposed by Curtright and Zachos [5].
\vskip .3cm
\par
Finally a few comments on the primary and descendant states. We define
the primary state corresponding to a primary field
$\Phi(w,\overline w)$ of weights $(h,\overline h)$ as
$$\vert h,\overline h>_{q}= \lim_{w,\overline w \to 0}
\Phi(w,\overline w)\vert
0>_{q}\eqno(39)$$
in close anology with standard case. In particular, for a
holomorphic field with weights $(h,0)$ the primary state $\vert
h>_{q}\equiv \vert h,0>_{q}$ can also be defined as
$$\vert h>_{q}=\phi_{-h}\vert 0>_{q}.\eqno(40)$$
Note that the modes $\lbrace \phi_{m}\rbrace$ for $~m\ge -h+1$ must
annhilate the "in" vacuum as a requirement for the regularity of
$\Phi(w)\vert 0>_{q}$ at $w=0$. Using this fact together with (16a)
and (37) we have
$$L_{n}\vert h>_{q}=0~~~~~~{\rm for}~~ n>0\eqno(41)$$
and
$$L_{0}\vert h>_{q}=q^2 [h]\vert h>_{q}.\eqno(42)$$
The $q$-descendant states are then constructed by subjecting the
primary states to operations of $L_{n}$'s for $n<0$:
$$\vert h;k_{1},k_{2}\ldots k_{m}>_{q}=L_{-k_{1}}L_{-k_{2}}\ldots
L_{-k_{m}}\vert h>_{q}.\eqno(43)$$
In passing, we would like to remark that it would also be
interesting to study the conformal properties of secondary fields
which give rise to the above $q$-descendant states. These together with
the primary fields would then constitute a basis for the study of
$q$-string theory.
\vfil
\eject
{\bf REFERENCES}
\ref 1 L.D. Faddev, in: Les Houches Lectures in 1982 (Elsevier,
Amsterdam, 1984);
\spa {~} P.P. Kulish and E.K. Sklyanin, Lecture Notes in Physics Vol.
151 (Springer, Berlin 1982).
\ref 2 V.G. Drinfeld, Proc. Intern. Congress of Mathematicians
(Berkley 1986) Vol. 1 pg 798;
\spa {~}M. Jimbo, Lett. Math. Phys.{\bf 11} (1986) 247.
\ref 3 E.K. Sklyanin Funct. Anal. Appl. {\bf 16} (1982) 262;
\spa {~}P.P.Kulish and N.Y. Reshetikhin, J. Sov. Math. {\bf23}
(1983) 2435.
\ref 4 A.J. Macfarlane, J. Phys. {\bf A22} (1989) 4581;
\spa {~}L.C. Biedenharn J. Phys. {\bf A22} (1989) L873.
\ref 5 T. Curtright and C. Zachos, Phys. Lett. {\bf B243} (1990) 237.
\ref 6 M. Chaichan, P. Kulish and J. Lukierski, Phys. Lett. {\bf B237}
 (1990) 401.
\ref 7 N. Aizawa and H. Sato Phys. Lett. {\bf B256} (1991) 185.
\ref 8 A.A. Belavin, A.M. Polyakov and A.B. Zamolodchikov, Nucl.
Phys. {\bf B241} (1984) 333.
\ref 9 M. Chaichan, A.P. Isaev, J. Lukierski, Z. Popowicz and P.
Pre\v snajder, Phys. Lett. {\bf B262} (1991) 32.
\ref {10} D. L\"ust and S. Theisen, Lecture Notes in Physics Vol
346 (Springer-Verlag, Berlin 1989).
\vfil\bye